\documentclass[twocolumn,showpacs,preprintnumbers,amsmath,amssymb]{revtex4}
\usepackage{graphicx,psfrag,bbm,latexsym,color,dcolumn,bm,dsfont,bbm,color,mathrsfs,bbold,latexsym,amsmath,amsfonts,amssymb}

\newcommand{\beq}{\\ \begin{equation}}
\newcommand{\eeq}{\end{equation} \\}

\newcommand{\beqa}{\begin{eqnarray}}
\newcommand{\eeqa}{\end{eqnarray}}

\newcommand{\bit}{\begin{itemize}}
\newcommand{\eit}{\end{itemize}}

\renewcommand{\S}{{\cal S}}

\newcommand{\tr}{{\rm Tr}}

\def\t{\tilde}

\def\Re{ \mathop{\rm Re} }

\newcommand{\ident}{\equiv}

\newcommand{\ebh}{e^{-\beta \cH}}

\newcommand{\cH}{\hat {\cal H}}

\newcommand{\TO}{\!\rightarrow\!}

\begin{document}

\title{Asymptotics of the Infrared}
\author{P.R. Crompton}
\affiliation{
Institut f\"ur Theoretische Physik, Universit\"at Leipzig, \\
Augustusplatz 10/11, D-04109 Leipzig, Germany.}
\vspace{0.2in}
\date{\today}
\email{crompton@itp.uni-leipzig.de}

\begin{abstract}
\vspace{0.2in}
{We follow recent formulations of dimensionally reduced loop operators for 
quantum field theories and exact representations of 
probabilistic lattice dynamics to identify a new scheme for the evaluation 
of partition function zeroes, allowing for the explicit analysis of 
quantum critical phenomena. This new approach gives partition function 
zeroes from a factored quantum loop operator basis and, as we show, 
constitutes an effective mapping of the renormalization group 
$\beta$-function onto the noncommuting local 
operator basis of a countably finite Hilbert space. The Vafa-Witten theorem 
for CP-violation and related complex action problems of Euclidean Field 
theories are discussed, following recent treatments, and are 
shown to be natural 
consequences of the analyticity of the limiting distribution of these zeroes, and 
properties of vacuum regimes governed by a dominant quantum fluctuation in 
the vicinity of a renormalization group equation fixed point in the infrared.}
\vspace{0.1in}
\end{abstract}


\maketitle

A classical spin system and its quantum system 
analogue apparently share many similarities at the level of the spin 
Hamiltonian. The difference between quantum and classical formulations is 
perhaps somewhat subtle at first glance. In the latter case the 
spin components are represented through the action of an operator basis, 
rather than purely orthogonal vector basis elements as is the case 
with the former. Although the 
dimensionality of a particular model is 
unaffected by this step of operator inclusion the partition function of 
the quantum system is 
formally defined as a trace over this operator basis acting on some 
suitably structured Hilbert space (whereas the classical 
partition function is strictly given simply through a summation 
over spin vector realisations). 
This inclusion therefore adds an extra dimension to evaluation of the quantum 
partition 
function that maybe then dealt with, or expressed, through a variety of 
scenarios relating to the applicability of dimensional reduction. This basis 
reformulation or inclusion of operator components can in fact be argued to 
be a form of second quantisation of a given classical model~\cite{Dtheory}.

We are ultimately interested here to derive an exact polynomial expansion for 
the partition function for a generic quantum system formed in such an operator 
basis, and to discuss the effect that quantum fluctuations have on 
frustrating this effort. The zeroes of this polynomial expansion then 
constitute the 
partition function zeroes whose limiting distribution we aim to now analyse 
following the arguments and reasoning presented in~\cite{lee+yang1}\cite{fisher}. 
However, the difference 
from these two cases we will argue (and would hope to stress as a central result) 
is that in general we identify the partition function of a generic quantum 
system is nonanalytic and unamenable to such expression. Our 
reasoning will follow much 
as for the general case of renormalization group equations of 
motion and the $\beta$-function, which may become singular prior to reaching its 
asymptotic value through infrared divergences~\cite{strong}. 
Crucially dependent on structuring the formal 
limits in the countably finite Hilbert space basis we will argue a 
limiting analyticity can be defined, however, using recent results and discussions of the 
structure of the poles of the inverse Laplace transform in the context of 
recent results on exact lattice probabilistic dynamics~\cite{Wa}\cite{Wb}. 

It may initially appear some what confused to attempt to define bounds for the 
analyticity of this expansion, or indeed of little practical relevance to 
order and arrange a form for an exact expansion that is valid essentially only in the 
vicinity of the continuum limit. This is certainly somewhat of 
an anathema to the scaling arguments of the renormalization group and partition function 
zeroes formulations where scale invariance is a central point. 
Our major thrust, however, will 
be to treat disconnected contributions to the partition function in a local 
basis, marking a certain departure from (quasi-classical) perturbative 
expansion. The noncommutativity of the operators for the quantum system 
force the reformulation in this regard. Diagrammatically 
disconnected contributions will be here suppressed through the limiting 
distributions of the basis operators formed on the Hilbert space, 
rather than through the tacit convergence of any effective series.
Akin to partition function zeroes formulation, singularities are expressed 
through the limiting asymptotic behaviour of poles as the continuum is 
approached, but here with an intermediate result closely 
connected with dimensional reduction that factorises a local symmetry 
preservation.
To facilitate this discussion we first review salient points in the 
derivation of the transfer matrix and associated Quantum Monte Carlo method 
for assigning exact probabilistic dynamics to a loop operator basis, although 
we would stress this is not primarily intended as a discussion to motivate 
specific numerical approaches. 

\section{Noncommutativity of the Hamiltonian Operator}

A number of algorithmic scenarios exist for the Monte Carlo simulation of 
quantum spin systems ~\cite{loop}\cite{factorc}\cite{factorb}\cite{beard}\cite{sse}. 
They were developed for the purpose of addressing the 
critical 
slowing down that affects the simulation of quantum systems at low 
temperatures. They differ largely from standard lattice gauge theory 
simulation methods, for example, in that in the main they are driven locally. 
Probabilistic Monte 
Carlo decisions are taken on a lattice site-by-site basis rather for the 
global 
lattice configuration as is usually the case with the latter~\cite{R}. Although, we 
will later comment on the relevance of the results to the complex action 
problem and recent lattice attempts at a general solution in this direction 
here algorithms for the numerical study of quantum 
spin systems serve as a means to introduce the formulation a 
complete-orthonormal operator basis acting on a Hilbert space in the 
context of exact lattice probabilistic dynamics. Following recent 
discussion on the limiting continuum distribution of a suitably arranged 
Markov process, we then argue this is amenable to analytic expression. 
Thus a means is identified to formulate 
partition function zeroes to express singular behavior in a quantum 
system for cases where otherwise the non-commutativity of the quantum 
operator will basis spoils this analyticity. This result is so expressed 
through an approach closely connected to dimensional reduction~\cite{Dtheory}. 
From this derivation it becomes then possible to the define bounds under 
which a generated Monte Carlo distribution becomes suitably close to 
the continuum 
result that the zeroes of the partition function can be meaningfully associated
with the limiting thermodynamic arguments for the singularities 
of \cite{lee+yang1}\cite{fisher}. Or more 
properly now their quantum system analogues, as we will argue from the 
results on the behaviour of the poles of the inverse Laplace transform of 
this expansion. In this sense, therefore, 
determining the statistical limitations of numerical Monte Carlo sampled 
loop operator distributions with relevance to this task remains an onerous 
and somewhat separate ongoing direction, although we return to discussion of 
the complex action problem in a later section. 

Proceeding with commenting on generic Quantum 
Monte Carlo approaches for quantum spin systems, the pedagogical 
example usually 
taken is the spin one half anti-ferromagnetic Heisenberg model 
($S\!\!=\!\!1/2$ AFM)\cite{loop}\cite{factorb}. The transfer 
matrix of this system contains the simplest off-diagonal singlet contribution 
possible to express a quantum interaction.
The spins defined $S_i=1/2 \,\,\,\sigma_i$ are defined for the model 
on each lattice 
site index $i$ of a $d$-dimensional hypercube, with a nearest-neighbour 
interaction and $\sigma_i$ being the Pauli matrices. In general, ie. for 
all but vanishing off-diagonal contributions of the 
suitably defined transfer matrix, the spin 
operators do not commute. If one were to wish to evaluate the Boltzmann 
factor exactly from the nearest-neighbour spin Hamiltonian thus defined on a 
finite lattice volume this would then present something of a problem, akin to 
the difficulties of representing Grassmann algebras in lattice gauge field 
theories. The combined idea of Suzuki and Trotter as a basis 
for numerical simulation was to 
partition the nearest-neighbour Hamiltonian operator $\cH=\sum_i H_i$. For 
the case of the $S\! =\! 1/2$ AFM the leading result would be then 
given for an odd-even site partitioning of the operators. 
\beq
\ebh \approx\prod_i\exp(-\Delta\tau \, H_i)+{\cal O}(\Delta\tau^2)
\eeq
The error in estimating the exponential is thus subdominant despite the 
noncommutativity of the nearest-neighbour spin operators that constitute the 
Hamiltonian $\cH$. This is 
dependant evidently on the value of $\Delta\tau$, as yet undefined. 
The partition function itself is defined as the trace over the basis on 
which the operators act, and thus the extent of the above product $i$ must 
clearly be somehow dependent on the lattice volume of 
the classical spin model on which the quantum system operators are 
introduced. For the $S\!\!=\!\!1/2$ AFM, as we have said, the leading 
result is given through odd-even partitioning. 
There would be no reason a priori, though, to define a particular 
lattice extent for the additional dimension introduced by the operators 
themselves. 
Indeed, only two points need be formally defined for the trace 
operation. If $\Delta\tau = \beta/ L_t$ is the defined as 
the lattice spacing in Euclidean time the only necessary points to preserve 
the measure are $\Delta\tau=0$ and $\Delta\tau=\beta$. However, we wish to 
discretise the Euclidean time direction, not simply for the purposes of 
numerical simulation but to now enable discussion of the recent limiting 
results of probabilistic dynamics on this Euclidean extent by first treating 
it as a countably finite space.
To properly define the partition function ${Z(\beta)}$, it is necessary to 
first note that the elements of the factorised product are the transfer 
matrices that act on the suitably arranged complete-orthonormal set of basis 
states. 

\beqa
 {Z(\beta)}& \equiv & {\rm Tr} ~\ebh
 \approx \sum_{\{S_{it}\}} \prod_{p}  W_p (\{S_p\}) \,\\
 W_p(S_p) & = & 
  \langle S_{i,t  } S_{i+1,t  } | e^{-\, \Delta\tau H_{i}} |
          S_{i,t+1} S_{i+1,t+1} \rangle \;
\eeqa
The 
summation of the weights $W_p$ for the transfer matrices above 
therefore runs over the spin configuration space of the 
lattice, 
as given through the state of the elementary plaquettes $p$. For the 
$S\! =\! 1/2$ AFM these are defined as the two-by-two elements of the 
hypercube defined over a unit spatial $i$ and Euclidean time $t$ extent. 
The weight $W_p$ associated with the of each plaquette are 
given by the matrix elements, and can be built up numerically through the 
probability density function distributions generated by a Monte Carlo 
procedure, as we will later discuss~\cite{beard}. The definition 
of the transfer matrix elements and their explicit relation to the 
noncommutativity of the spin operators will also be postponed to later 
section, whilst we consider an alternate quasi-classical approach~\cite{sse}\cite{cumulant}.

The reverse of the above product expansion is of course possible. Rather 
than proceeding from a local, and 
inherently limiting-case factorisation perspective, one can aim to treat the 
statistical fluctuations of the noncommuting operator basis in a 
global manner, using the ideas of stochastic field theory.
Statistical cumulants $\langle H_{1}\ldots H_{N} \rangle^{c}$ 
can be defined on the operator basis through the 
differentiation of some suitable generating function, 
\beqa
\left.\langle H_{1}\ldots H_{N} \rangle^{c}\right. \! & \! = &
\nonumber
\\
\!\!\!\!\!\!\!\!\frac{\partial}{\partial
\lambda_1 }\cdots & & \!\!\!\!\!\!\!\!\! \frac {\partial}{\partial
\lambda_N} 
\ln \left\langle  e^{\lambda_1 H_1} \ldots \, e^{\lambda_N H_N} 
 \rangle~\right|_{\lambda_{1}=\ldots=\lambda_{N}=0}. \,\,\,\,\,\,\,
\eeqa
It 
is important to note that no assumption is being made 
for the distribution of the operators on the space in this. If this were the 
case one would arrive at a standard time-ordered path integral formalism 
for quantum field theory generating functionals, albeit 
on a countably finite space (a Fock space definition), 
rather than this statistical 
result which is simply for functions rather than functionals. That the
time-ordered result is not guaranteed is 
somehow obvious since the operators in the cumulant definition are as yet 
rather general and undefined. 
Their action could be highly nonlocal or could indeed be incomplete in 
some sense, in which case the dynamics would be beset by singularities. 
That the action of a proper 
Hamiltonian necessarily yields a time-ordered product is somehow less 
obvious, but will we again postpone a fuller discussion of issues of locality 
and noncommutativity to the section that follows, where the role of a Markov 
process in defining exact probabilistic dynamics for the lattice is made 
evident.

Following this analogy, though,  
the expectation value of the product defined without the logarithm in the 
righthand side would correspond to analogue of the mixture of disconnected and 
connected $n$-point functions one obtains from the action of the generating 
functional in the field theory analogue.
Again, without assuming form for the distributions one can take the 
continuum limit of the operator space $N\rightarrow \infty$ for the above 
definition of cumulants. This then yields a quasi-classical 
effective Hamiltonian $\cH_{\rm c}$ from equating the cumulant definition 
to an exact series expansion for an exponential in this limit. This appears 
evidently simpler to evaluate, being reduced to a quasi-classical formulation 
and therefore amenable to a summation over vector realisations. 
\beq
Z = \int \!\!{\bf n}\, \langle {\bf n}| \exp (- \beta \cH) |
{\bf n}\rangle
\eeq
\vspace{-0.45in}
\beq
\exp[-{\beta\cH_{\rm c}}({\bf n},\beta)] \equiv 
 \langle {\bf n}| \exp (- \beta \cH) |{\bf n}\rangle
\eeq
Caution is in order however as one must now expand in the matrix elements of 
powers of the full Hamiltonian $\cH$, for a practical numerical approach to 
evaluate the quasi-classical Hamiltonian $\cH_{\rm c}$. 

\beqa
 {\beta\cH_{\rm c} }({\bf n},\beta)  =  \, 
  \langle {\bf n}| \, \cH \, |{\bf n}\rangle^c 
 - \frac{\beta}{2!} & & \!\!\!\!\!\!\!
\langle {\bf n}| \, \cH^2 \, |{\bf n}\rangle^c +
 \nonumber \\ & & \!\!\!\!\!
\frac{\beta^2}{3!} \langle {\bf n}| \, \cH^3 \, |{\bf n}\rangle^c + \ldots
\eeqa
As was case with 
the Trotter-Suzuki approach the noncommutativity of the basis cannot be 
entirely circumvented. This expansion may appear somewhat straight forward, 
but $\cH_{\rm c}$
itself is strictly defined only through the cumulants (connected components), 
rather than the statistical expectations (connected and disconnected 
components). A numerical evaluation of the quasi-classical system following 
similar truncation approximations to the Trotter-Suzuki form therefore suffers 
from three sources of systematic error. The expansion itself may diverge 
through either the value of $\beta$ or the level-spacing diverging as a 
function of the expansion, and the stochastic expansion terms themselves 
contain the statistical analogue of disconnected contributions to the 
quasi-classical partition function in all but the continuum limit. 
Without unduly pessimistic focus on the numerical prospects, both of the two 
parallel 
approaches fall on this issue of noncommutativity. Yet at the same time, 
this comparison as the continuum limits are separately approached is revealing.

To make a further association with issues of dimensional reduction in gauge 
field theories we can aim to relate quasi-classical statistical and 
functional terminologies for effective actions.
At small inverse temperature $\beta$, the interaction coupling of gauge 
fields becomes small through asymptotic freedom and perturbative expansion 
is justified~\cite{DimReda}\cite{DimRedb}. 
For length scales much larger than those of the Euclidean time extent it 
also seems reasonable to integrate out the nonstatic (time-dependent) 
modes of the gauge fields. This gives the leading behavior at small 
$\beta$, but it is important that the perturbative expansion of this
dimensionally reduced model does not have divergences in the infrared. 
The nonstatic modes are suppressed and renormalize the expansion 
terms of the new effective theory.
To develop the above statistical association, 
the expansion is again valid if the perturbative coupling is not so large as to cause 
divergence. By way of contrast no explicit assumption is made 
about the locality of the expansion
in the statistical case and so it is not renormalized.
The dimensional reduction of the field theory 
here forms the functional analogue of the stochastic 
approximation. The functional approach becomes again invalid if 
the contribution of higher order $n$-point functions 
(either connected or disconnected) becomes dominant. 

It is essentially 
the locality of the 
expansion of the effective action that crucially determines the response to 
infrared divergences and so we seek a different approach as a basis for study. 
By following the Trotter-Suzuki decomposition, 
as opposed to the effective quasi-classical Hamiltonian formalism, we work 
in a nonlocal framework without implicit effective series truncation. 
Lee and Yang, in the context of classical spin systems defined a limiting 
analyticity for the partition function such that thermodynamic singularities 
were only strictly expressed through the limiting distributions of the 
zeroes of polynomials so constructed~\cite{lee+yang1}. 
We we will argue that this specific 
analyticity property of the partition function is a nontrivial consequence 
of the renormalization group approach in the quantum case by working over this 
extended phase space. 
By addressing the form of the limiting distribution taken by noncommuting 
operators on the basis space we will seek to similarly define a limiting 
analyticity for the infrared divergences connected with the renormalization 
group equation fixed points of quantum systems through this 
factorisation approach.

\section{Loop-Cluster Operators}
Before proceeding to discussion of 
the partition function polynomial form for the quantum case we continue first 
outlining the relationship between the factorised 
Hamiltonian operators and their distributions on a Hilbert space. This 
is the relationship defined through the transfer matrix elements of these 
operators. We motivate these now necessary definitions through 
part-review of numerical Quantum Monte Carlo loop-cluster methods. 
Numerical motivation perhaps seems an unnecessarily heuristic path to take 
here for the proposing of an exact continuum form for 
the generic partition function of a quantum system. 
To reiterate, indeed the stated aim is not to define 
numerical grounds for simulation. By constructing, or reviewing, the Markov 
process definition we will simply articulate the complete-orthonormal nature 
of the basis formed by the elementary plaquettes defined over the spatial and 
temporal sites, and exact probabilistic dynamics defined thereon. 

Again whilst a factorisation property is clearly a key element in 
properly defining the zeroes of a polynomial, the direct connection with a 
complete-orthonormal basis for a Hilbert space and partition function zeroes 
of a generic quantum system would perhaps appear obtuse at this point.
However, mirroring Lee and Yang's discussion 
we associate the analyticity of an 
exact expansion with identifying the continuous distribution or functional 
assignment of solutions to the generating functions of said expansion. 
That each expansion coefficient in the polynomial expansion can represent a 
well-defined integral equation is actually a necessary starting point of 
Lee and Yang's discussion \cite{lee+yang1}. We therefore do this explicitly here by 
considering the action of taking the vanishing Euclidean time lattice spacing 
limit for the loop operators on a finite spatial lattice extent. The 
construction of the quantum system analogue of Lee and Yang's integral 
assignment is then given through this action, implicit to the loop-cluster 
method definition.

The loop-cluster method for quantum systems is similar in form to the 
that of the Swendsen-Wang cluster algorithm of classical statistical 
mechanics~\cite{S+W}\cite{loop}\cite{factorb}. 
The partition function is defined over a set of spin configurations  $\{\S\}$ 
of the lattice, and a set of graphs $\{G\}$ linking these spin states, 
defining an extended phase space. For the Monte Carlo process statistical 
weight functions $W(\S,G)$ must be assigned for each spin configuration of a 
lattice ensemble, such that now all elements of the extended phase space are 
defined as either being connected or disconnected. Given some configuration 
of the 
spins $S$ of a lattice each bond linking spins is separately considered 
and assigned such a status based on the value of $W(\S,G)$, determined 
through the local spin configuration of $S$. This then 
implicitly defines a path, or graph 
subset $G$, through the volume of the particular configuration. 
The total weight of a 
given configuration $W(S)$ in the lattice ensemble is given by the sum over 
graph weights.
\beq
  \sum_G   W(\S,G) =  W(\S) \,\,\,\,\,\,\,\,\,\, p(\, \S \,\TO\, (\S,G)\,)  
\;=\; \frac{W(\S,G)}{W(\S)} \;,
\eeq
where $p(\, \S \,\TO\, (\S,G)\,)$ is probability of a obtaining a particular 
graph within a lattice configuration. This then
implies both that $W(\S)\ne0$ and $W(\S,G)\ne0$. 
The process is otherwise nonergodic and there is some absolute preferred 
configuration state of the volume that once reached cannot be left in an 
exact statistical sense. We are interested to treat singularities, of course, 
but do not anticipate divergences such as these on a finite 
lattice for essentially geometric reasons - without, though, wishing to 
necessarily restrict the spaces of interest at present. 
The loop-cluster algorithm itself is clearly of 
more general construction than its applicability to quantum systems, but the 
Hamiltonian operation we are interested in, acting locally on suitably 
defined plaquettes, can satisfy this latter Monte Carlo probabilities 
detailed balance condition.

Returning to the importance of locality. Considering the pedagogical 
example of the $S=1/2$ AFM according to the above 
definition of a graph a single 
spin cannot be inverted alone on an elementary plaquette to create a valid 
change between successive lattice configurations. This follows from the 
implicit local conservation of $S$. The graphs so defined thus  
map relative change across the extended volume. 
In fact, 
demanding on a given a graph that the form of $W^{plaq}(\S,G)$ does not 
change at all between 
consecutive lattice configurations in an ensemble 
then implies that all allowed plaquette 
changes match up between neighbouring plaquettes 
across the volume. 

An extensive literature exists on the definition of suitable definition of 
transfer matrices for quantum systems and the loop-cluster method, but  
to clarify further for the pedagogical example the here plaquette has eight 
spin configurations, $S_p = i^\pm$, of the four site vertices, $i$ \cite{loop}. 
Each transition $i\leftrightarrow j$ defines a respective transfer matrix 
element $w^{ij}\ident w^{ji}$, and corresponds in this case to inverting 
either two or all four of the spins of a plaquette. We can then therefore 
think to write the partition function for this, or a generic quantum system, 
in terms of an equivalent bond or loop operator $h_b$ basis defined 
only in terms of the pertinent transfer matrix 
elements~\cite{loop}\cite{factora}\cite{poissona}
. The pertinent 
transfer matrix elements will be those are nonzero. The diagonal elements of 
these loop operators for the pedagogical case will correspond to four 
inversions and comprise the identity operator, and the self-adjoint 
off-diagonal remainder will be given uniquely through the graphs determined 
for each configuration (or rather their ensemble-average). This all then 
essentially represents an effort to define the pseudo-reduction step on a 
finite basis, mapping the discrete distribution of graphs to the limit of 
vanishing Euclidean time lattice spacing where a continuous distribution 
is recovered. The partition function in terms of the loop operators 
$\hat{h}_b$ is, 

\beqa
  \tr\,\, e^{-\beta \cH}
  &= & e^{\beta \sum_b J} \lim_{\Delta t\to 0} 
     \left({\prod}_b e^{(-J +J h_b) \Delta t}\right)^{\beta/\Delta t}\\
  &= & e^{\beta \sum_b J }\int\rho(d\omega)\prod^* h_b ~,
\eeqa
where the Hamiltonian is rewritten over the bonds $b$ as $\cH = - \sum_b J\, 
h_b$ and $J$ is the nearest neighbour coupling. Here the 
$\prod^* h_b$ is a time ordered product of the loop operators with 
$\rho(d\omega)$ a Poissonian probability measure of the connection status 
recovered by taking the limit of vanishing Euclidean time lattice spacing. 
That surprisingly this relatively general construction of the loop operator 
transfer matrices yields a precise continuous distribution in the limiting 
step follows directly from the locality property.
It is important to note as well that no noncommutativity property is invoked 
to obtain the Poissonian distribution on the basis elements of the now 
countably finite spatial element space. One could have equally well 
ignore the off-diagonal elements altogether and used the approach to map 
a classical system into another classical system, since after all the method 
is classical in origin. 

The advantage over the quasi-classical series expansion 
approach here is that the reduced factorised basis elements are explicitly 
highly nonlocal in character whilst locally preserving $S$ exactly. 
The central assumption, however, is again that of the quasi-classical 
expansion : the statistical averages only become interchangeable with 
those of the thermodynamic or dynamic averages in the infinite volume limit. 
Here the definition is somehow less ad hoc and the treatment of disconnected 
contributions less tacit, since locality is enforced site-by-site, which 
allows us define a polynomial with a slightly different relation to the poles
of the partition function. 
By construction, the basis of plaquettes is complete, and orthonormal and more 
importantly the weight functions associated to graphs are unique and 
independent of the location within the Markov chain or statistical ensemble. 
Although 
$S$ is locally preserved the elements of the extended space 
reached by a finite Markov process can be necessarily incomplete and so 
functional solution is not necessarily admitted. We could be trite and label 
this the ambiguities of numerical determination, but the issue of whether or 
not a discrete extended basis can form a covering of the target space is a 
perhaps a more interesting question which we will return to shortly. 

The extent to which the above 
partition function corresponds with the Nambu definition of a generalised 
Hamiltonian
(in the operation of elements on a Poisson manifold \cite{nambu}), and could perhaps then 
therefore be 
now further generalised for non-Euclidean metrics, is certainly an interesting 
question but not one we will aim to further address here.
We wish to now investigate the limiting 
analyticity of this factorised reduced expansion, noticing both that the 
probability distributions of the operators become exact and continuous 
in the limit of vanishing Euclidean time lattice spacing and that a 
product of two-by-two self-adjoint operators (for the $S=1/2$ AFM case, and of 
generic form) is particularly simple to diagonalise. We have now identified 
the continuous distributions associated with factored product which we are 
aiming to now diagonalise, in keeping with Lee and Yang's motivations, as a 
first step towards establishing the analyticity properties of the eigenvalues 
and characteristic polynomial associated with this operator product.

\section{Partition Function Zeroes Polynomial}

We have stressed, perhaps seemingly unnecessarily, that the 
partition function zeroes for a generic quantum system are somehow 
different than those introduced by Lee, Yang and Fisher 
which we interpret now as classical results. 
By considering the action of loop operator 
elements on an 
extended phase space, defined through a complete-orthonormal local basis, we 
have discussed in the previous section 
the identification of an exact factorised 
operator product on a countably finite Hilbert space with an associated 
continuous time-ordered distribution structuring. This 
is a limiting result, but as we have noted the finite basis loop operators 
by construction are trivially diagonalisable in general. 
The new step we now take is to note that there is a necessarily a difference 
between diagonalising the elements of a direct product and determining the  
characteristic equation of that product. Herein, we believe, lies the 
divide in the arguments concerning deriving an exact analytic form for the 
partition functions of the respective quantum and classical systems. 

If the distribution obtained 
through projecting the graph results onto the spatial extent of the lattice 
is discrete, rather than continuous, in the quantum case some 
of the loop operator elements may be defined only on a subspace of the 
extended space. The transfer matrix allows for the lattice to be 
entirely comprised of disconnected contributions, of course, but this 
singularity is under defined in the discrete quantum case. 
The eigenvalues of the factored 
product are therefore not necessarily analytically 
connected on the finite Hilbert 
space defined over the spatial elements with a discrete 
statistical distribution, ie. at finite Euclidean time lattice spacing.
Conversely with what we here 
deem the classical result (if this pictured simply as a limiting case of 
commuting operators) all elements are necessarily connected through the 
diagonality of all the terms of the factored product. 
There may be still disconnected contributions to 
the partition function but the space of the partition 
function itself is not now under defined as a consequence, since the 
plaquettes are constructed by definition to preserve a local symmetry. 
Up to  
singularities in the continuum limit for the Hilbert space, as defined over 
the loop operator basis of the lattice spatial extent, we can thus be 
assured that the 
classical case is always defined whereas in the general 
quantum case is not. At least, to be clear, if we are now aiming to treat 
the continuum limit of the eigenvalues of our product as the polynomial 
zeroes of the partition function of Lee, Yang and Fisher we must treat the 
spatial and Euclidean time extents separately in our analyticity arguments. 
This follows essentially as a direct consequence of the noncommutativity of 
the operator elements of the Hamiltonian in the quantum case.

In the quasi-classical series expansion formalism we would now have difficulty 
proceeding with our discussion as a basis for practical calculations 
because the basis operations in the 
extended phase are nonlocal. This means that we could only recover both 
of our limits simultaneously : the discrete statistical distribution being 
built up step-by-step is that for the entire volume, rather than 
for the elementary plaquettes alone. Conversely, the factored 
product elements we will now focus on are being defined towards the limit of 
vanishing Euclidean time lattice spacing, thus giving an analyticity ahead of 
the continuum limit for the spatial extent through a preserved local 
symmetry. It is this effective treatment of the disconnected contributions 
that allows us to proceed in the quantum case with the factorised product approach.

What we want to now consider, therefore, is if the eigenvalue spectra, 
assumed now complete and covering for the extended space, has a 
meaningful connection with the poles of the free energy density of the 
partition function and if these poles then demonstrate the lattice volume scaling 
properties associated with partition function zeroes analysis of Lee, Yang 
and Fisher. We do this now by noting 
recent results on the structure of the inverse Laplace transform poles of 
the transfer matrix elements in the context of discussions on exact 
probabilistic lattice dynamics. 
To follow this recent discussion, and deduce our zeroes continuum limit 
behaviour, we can rewrite the loop operator expression determined for the 
partition function through the notation of~\cite{Wa}\cite{Wb}. In this case 
multiplying by the 
infinitesimal Poissonian probability elements in the Euclidean time 
intervals $[t_1,t_1+dt_1], \ldots, [t_N,t_N+dt_N]$ the off-diagonal operator matrix 
elements $\mathcal{W}_{N}(\beta)$ of $h_b$ are now expressed as,
\beqa
&&\mathcal{W}_{N}(\beta;V_0,V_1,\dots,V_N) = \nonumber\\&&
\epsilon^N \,\, \left\{ \prod_{i=1}^{N} 
\,\, \int_{0}^{\beta} dt_i \,\, \theta(t_{i}-t_{i-1}) \,\, 
\exp\Delta_{i}t_{i}) \right\} \,\,e^{-V_{N}\beta}\,\,\,\,\,\, 
\eeqa
with $\Delta_{i}=V_{i}-V_{i-1}$ the potential energy difference 
between the sites at $i$ and $i\!-\!1$, and $\epsilon$ 
some arbitrary energy scale for the purposes of normalisation. 
In the previous section we showed that in the limit of vanishing 
Euclidean time lattice spacing we recover a time-ordered 
Poisson distribution for the elements of the factored spatial basis.
Here this is the initial assumption for the product : an element remains in a 
given basis state until it undergoes a Poissonian distributed event. We can 
thus consider this the generic form for the loop operator elements 
in this limit. 
Here we are again expanding the product over 
the basis of states $N$, the extent of 
which corresponds to the spatial volume, with the integrand here 
corresponding to the time-ordered limit of the graph distributions. In the 
discrete case the graph paths can propagate throughout the configuration 
volume, and equally here the product elements are in causal connection.
The above element maps between consecutive spatial sites $i$, with all four 
corners of the elementary plaquette considered by definition, and again $S$ 
is locally conserved. 
The explicit difference, however, between this weight form 
$\mathcal{W}_{N}(\beta)$ and $h_b$ lies 
in the expression of the specific matrix element contributions. 
The Heaviside step function being unidirectional over the product indexing 
here implies that the mapping of $\mathcal{W}_{N}(\beta)$ defines only one 
element of the transfer matrix. The time-ordering is here implicit to the 
potential energy values recovered from the 
graph determination process, whereas the ordering of the product indexing 
simply labels the element within the elementary plaquette. 
Since this latter ordering maps between successive site indices $i$ we can 
associate $\mathcal{W}_{N}(\beta)$ with the off-diagonal contribution to the 
loop operator $h_b$. 
Why we believe it relevant to consider just this subset of $h_b$ is that for 
the generic form of $h_b$ for a quantum system, as we have argued (somewhat tacitly)
in the previous section, the 
diagonal elements form the identity operator. The zeroes of the operator product are 
thus characterised by the values of the self-adjoint off-diagonal 
contributions. 

Notice again that the above expression is a result for an exactly 
ergodic 
Markov process, and so we must have achieved this statistical limit of 
vanishing Euclidean time lattice spacing for the above relation to be valid. 
We have argued, somewhat heuristically, 
that this limit is reached by generating a sufficient 
number of the formally defined graphs such that these graphs then recover 
a suitably complete 
basis for the extended space. Since the transfer matrix elements are as well 
defined 
over the ensemble rather than for a single lattice configuration, 
similarly we will be interested in the zeroes of the ensemble-averaged 
characteristic polynomial rather than those of a single configuration for 
a finite Euclidean time extent.
The above relation for the weight form $\mathcal{W}_{N}(\beta)$ is of course 
given in terms of potential energies rather than the polynomial zeroes of the 
loop operator product, but the analyticity of the zeroes and poles are in 
fact intimately connected through the properties of the inverse Laplace 
transform of $\mathcal{W}_{N}(\beta)$, which we will now consider 
following~\cite{Wa}\cite{Wb}. 
Differentiating with respect to $\beta$ and substituting for the Laplace 
transform $\widetilde{\mathcal{W}}_{N}(z)$ of $\mathcal{W}_{N}(\beta)$ 
reduces the weight form to the following recursive equation,  
\beq
\widetilde{\mathcal{W}}_{N}(\beta) =  
\int_{0}^{\infty}d\beta \,e^{-z\beta} \, {\mathcal{W}}_{N}(\beta)\,, 
\quad z \in \mathbb{C}
\eeq
\vspace{-0.2in}
\beqa
\partial_{\beta} \mathcal{W}_{N}(\beta) & = & \epsilon 
\mathcal{W}_{N-1}(\beta)  -V_{N}\mathcal{W}_{N}(\beta)\,, \\
\widetilde{\mathcal{W}}_{N}(z) & = & 
\epsilon (z+V_{N})^{-1} \widetilde{\mathcal{W}}_{N-1}(z) \,,\\
& = & \epsilon^N\prod_{i=0}^{N} \frac{1}{z+V_{i}}\,. 
\eeqa
The motivation for introducing these new relations is essentially to quantify 
the asymptotic scaling behaviour of the expansion via discussion of saddle 
point solutions. The above poles of the Laplace transform are by
definition confined to be in the real segment 
$[-V_{\mathrm{max}},-V_{\mathrm{min}}]$ , with 
$V_{\mathrm{max}}$ and $V_{\mathrm{min}}$ respectively the maximum and minimum 
elements of the set of potential energies $\{V_i\}$.
It follows from the definition of the Laplace inverse transformation 
that its integration contour $\Gamma$ is any line parallel to the 
imaginary axis contained in the analyticity domain of the Laplace transform, 
ie. $\Re z > -V_{\mathrm{min}}$.
Applying a saddle point approach to the expansion it can be shown that 
the first $N-1$ solutions lie in the range 
$-V_{\mathrm{max}} < \Re z_0 <-V_{\mathrm{min}}$ with the last solution lying 
such that $\Re z_0 > -V_{\mathrm{min}}$ , where $z_0$ are the locations of the 
saddle points and the ordering of solutions is in terms of increasing size. 
The restrictions on the poles come from quite general considerations, the 
only real condition here being that $\beta$ is necessarily 
positive~\cite{laplace2}\cite{laplace3}.

The whole point or purpose of the saddle point analysis here is that its solution is 
independent of the path distributions of the graphs but rather 
dependent on the 
local properties of the graphs as given through the potential energy 
differences. From this consideration then 
 the integration contour can be deformed to a new parallel 
one, although we must know by definition that the path mapped by the weight is analytic 
along the contour $\Gamma$ and where the isolated singularities exist. 
One could therefore then in principle equally well treat the general case of 
complex-valued potential 
energy differences, which would be similarly defined since $\Gamma$ would have no 
explicit singularities from the local properties of the graphs by 
construction. Generally, of course, there is no reason to suppose that 
the poles have to be uniquely defined, and could be given only up to a 
branch cut, but this is not a problem with the asymptotic form since in the asymptotic 
limit this choice is naturally resolved. We should note that through the above 
definitions and those of the previous section 
we are at the same time uniquely establishing the 
residue properties of the weight function and thus the 
existence of the zeroes of the loop operator factored polynomial.

The last saddle point solution alone is compatible with the restrictions 
on the 
integration contour $\Gamma$, and can be further shown to be bounded from 
below by $\Re z_0 > -V_{\mathrm{min}} +N_{V_{\mathrm{min}}}/\beta$~\cite{Wa}. 
Thus for 
$\beta\to\infty$ with $N_{V_{\mathrm{min}}} \sim N \sim \beta$ the distance 
between the saddle point and the closest pole remains finite, and the zero 
temperature continuum limit is well-defined. The distribution of the 
properly defined complete zeroes now necessarily have some limiting 
distribution with relevance to the continuum from the asymptotic behaviour of 
these poles.
To be clear, the partition function is expanded in whatever 
constant diagonal prefactor is introduced in $h_b$. It is proportional to 
$\beta$ for the pedagogical $S=1/2$ AFM example and therefore the analysis is 
specifically of interest for (although not restricted to) 
the case of quantum fluctuations at low temperatures. 
Strictly, as we have seen, only one of the poles lies in the analyticity 
domain of the polynomial. This does not imply that the characteristic 
polynomial is undefined, but rather that we only understand the asymptotic 
behaviour of the weight function in the vicinity of this one pole. There is a circle 
theorem by Lee and Yang which we will now discuss which relates to the 
limiting behaviour of branch cuts, which we can perhaps envisage now  
being applicable in at least one instance of the zero temperature limit. 
To reiterate, the noncommuting nature of the quantum system Hamiltonian 
has forced us to consider a limiting Euclidean time spacing distribution
defined over the spatial basis of our lattice, for which at least one of the 
zeroes of the polynomial by construction has a well defined continuum limit.   

\section{Scaling of Partition Function Zeroes}
We have considered how a partition function can be defined over an 
extended phase space for a quantum system through a loop operator formalism, 
and also how the limit of vanishing Euclidean time lattice spacing is related 
to the analyticity properties of the characteristic 
polynomial of this factored product and its pole structure. Although the recurrence 
relations defined for the above Laplace transforms of the operator elements have an
explicit dependence on the spatial lattice volume $N$, we must further explore this 
connection to firmly establish the zeroes content regarding the singularities of the 
partition function as the limit $N\!\rightarrow\!\infty$ is approached. The starting 
point of Lee and Yang's arguments is essentially that a valid zeroes polynomial 
form exists, from which finite volume scaling discussion can be then 
made~\cite{lee+yang1}. 
We have 
now argued that the analyticity of factorisation property of a quantum partition 
function is in general not possible without first approaching an asymptotic scaling 
regime : a far weaker precondition for this zeroes form. Comparing with
the constraints we have thus far defined on the factored product eigenvalues for the 
quantum system we now contrast with the two central criterion of 
Lee and Yang's discussion for defining valid thermodynamic functions from the 
partition function in the limit $N\!\rightarrow\!\infty$ of the spatial lattice 
volume~\cite{lee+yang1}. 

The first of these criterion is that the limiting value of  as 
 the pressure of the system $P$, is independent of the shape of 
$N$ and is a continuous monotonically increasing function of the expansion parameter 
of the suitably factored partition function. 
\beq
P = \lim_{N \to \infty} \,\, \frac{1}{N} \,\, {\rm{log}}(Z)
\eeq
Since the action of the factorisation in the quantum case is by definition 
local and ergodic it follows that the weights are obtained recursively, as we have 
seen. The limit 
$(N+M)\!\rightarrow\!\infty$ of the difference of the log of two consecutive 
intensive weights of integer index $N$ and $(N+M)$ is thus zero. 
Establishing this property one can then consider 
the volume as consisting of a finite subset of smaller volume elements, with 
interactions across the boundaries between subsets for which the free energy of this 
interface tension does not grow faster than the surface area. The above result thus 
implies that the limit of thermodynamic functions, 
including the pressure, is always finite. 
Monotonicity and the continuous nature of the functions follows from the similar 
behaviour of the derivatives of the partition function, as is seen explicitly 
through the properties of the inverse Laplace transform. Thus the distribution of 
the zeroes necessarily gives all the analytic behaviour of thermodynamic functions 
in the expansion parameter plane as $N\!\rightarrow\!\infty$. 

The second of Lee and Yang's 
criterion is that the limiting value of the density of the system
$\rho$, is an increasing function of the expansion parameter of the suitably factored 
partition function but does not necessarily take a defined limit for all 
expansion parameter values, $\beta$. 
\beq
\rho = \lim_{N \to \infty} \, \frac{\partial}{\partial\, 
log(\beta)} \,\, \frac{1}{N}\,\, {\rm{log}}(Z)
\eeq
Again this is very much an implicit result for the factored product 
of the quantum system Laplace transform and the discussions of the previous section
. The criterion implies that if a 
region of the positive real axis of $\beta$ is free of zeroes of the partition function 
as $N\!\rightarrow\!\infty$ then the system is here in a single phase. The proof 
of this follows from noticing that the coefficients of the factored product expansion 
are necessarily positive, which is a fairly general consequence of the operators being 
self-adjoint in the quantum case.

Lee and Yangs first discussion with their new result was essentially 
to notice that there was 
necessarily a problem in discussing thermodynamic functions with solely the 
$N\!\rightarrow\!\infty$ limit of the partition function polynomial expansion 
coefficients : this leads to the result of undefined phases. It was not possible 
to discuss thermodynamic singularities directly in the asymptotic limit, rather the 
convergence to that limit was the important consideration. They reviewed an 
earlier treatment of Mayer in which they showed it becomes impossible to 
analytically continue the series expansion thus formed between phases. This is 
unlike of course, by construction, the above implicit spatial lattice volume 
dependencies of $P$ and $\rho$ given from the properties of the Laplace 
transforms of the operator elements that do yield well defined thermodynamic functions. 
The difference lies in the differing properties of a series and polynomial.
The series expansion values do not yield 
a density dependence to pressure for specific volumes below the point at 
which the zeroes approach the real $\beta$ axis (the phase transition point). 
This is implicit since the continuum values have obviously no spatial lattice volume 
dependence.
One can therefore argue from this that not only is the partition function zeroes form 
for the quantum system compatible with Lee and Yang's discussion, but also that it is 
therefore in some sense a necessary precondition for all partition function zeroes 
expansions, classical and quantum. To expand : we have laboured to ensure the factored 
product is meromorphic, singular only at the poles, which are themselves in a 
one-to-one correspondence with the zeroes. 
A loop operator picture defined over an extended phase space must be 
therefore factorisable for all the thermodynamic functions to be able to defined for 
a system, since it is only the existence of the limit of 
vanishing Euclidean time lattice spacing that ensures that the polynomial expansion 
zeroes, and by extension the expansion coefficients, have a valid asymptotic form.

We can generalise this discussion further to consider the implications for complex 
operator elements and the introduction of symmetry breaking source terms defined 
with relation to the limit $N\!\rightarrow\!\infty$. 
Partition function zeroes have been recently used to discuss the 
Vafa-Witten theorem and complex action problems for quantum systems~\cite{vafa}. 
To summarise 
this discussion : the Vafa-Witten result is that parity and CT cannot be 
spontaneously broken in vector-like theories such as QCD. The argument 
follows from two points ; any arbitrary local hermitian order parameter 
for parity $X$, constructed from Bose fields is proportional to an odd power of the 
totally antisymmetric tensor, and the free energy 
density is well-defined in the presence of a symmetry breaking source $\theta X$, 
with $\theta$ some real constant. These two conditions jointly then imply that 
the symmetry breaking source term be a phase in the integrand of the 
Euclidean partition function. The explicit connection of the theorem 
with partition function zeroes is then made by considering $p(\t{X} ,N)$ 
the probability density function of 
$\t{X}$, where $\t{X}$ expresses $X$ extensively.
Assuming parity to be spontaneously broken $p(\t{X} ,N)$ can then be argued to be 
taken as 
the sum of two delta functions in the limit $N\!\rightarrow\!\infty$, centred on 
$\pm a$. The relevant zeroes of the partition function in $\theta$ are thus obtained 
as the solutions of, 
\beq
\cot (\theta Na)=
{{\int^{\infty}_{0}  d\t{X} \,\, p(\t{X},N) \,\sin\!\left(\theta N(\t{X}-a)\right)
}\over
{\int^{\infty}_{0}  d\t{X} \,\, p(\t{X},N) \,\cos\!\left(\theta N(\t{X}-a)\right)
}}\,\,.
\eeq
This has an infinite number of solutions approaching the origin, and so
 the Vafa-Witten theorem is thus seen to be upheld since the second tenet of 
free energy is then undefined. 
In fact this is a general consequence of the properties of the Laplace transforms of 
complex-valued integrands. The asymptotic convergence, as a function of $\theta$, 
is in general not defined for complex-valued functions given for the exponent : the 
oscillatory phase means the integration contour $\Gamma$ 
is ill-defined. Conversely, as we 
have argued, following Lee and Yang's discussion of Mayer's series treatment it is 
possible to consider the symmetry breaking consequences of a complex-valued system 
if the expansion coefficients are expressed intensively. Namely, if we have a 
contribution to the partition function that is highly nonlocal but preserves an 
exact local symmetry the asymptotic convergence properties of the partition function 
are defined, as we have seen explicitly with the above loop operator formalism. 

We can understand in a very simple way how singularities of the partition function 
arise as $N\!\rightarrow\!\infty$ from our pedagogical loop operator example. 
Consider a self adjoint matrix with real elements of values between zero and one and 
the product of all elements in the upper triangular region such that these give 
recursively the characteristic polynomial expansion coefficients to successive 
order~\cite{lee+yang2}. 
It can be shown by induction that if these elements have this probabilistic 
interpretation then necessarily the zeroes of the characteristic polynomial are 
bounded on the unit circle. In the case of Lee and Yang's discussion for 
the Ising model with external 
field we can then identify the field associated with generating the off diagonal 
transfer matrix elements and a symmetry breaking nonzero source value. 
Although here there is 
no such association we can still understand the singularities of the spatial extent 
of the partition function in the same way, and therefore can 
similarly associate a critical value to our factored product expansion parameter 
with a valid asymptotic form. Generalising for complex valued operator elements we see 
that the critical value of coupling is still recovered through a similar 
boundedness induction argument as the operators are necessarily hermitian in this case. 
We have in the loop operator introduction discussion and subsequent factorised 
product discussion essentially defined a source term such that it is only recovered 
in the asymptotic limit, thus obviating 
the ambiguity of contour definition for the Laplace transform for the complex case. 

Our initial motivations included a comparison of quasi-classical 
and factored product expansions for the treatment of quantum systems. It is perhaps 
now more evident why we have favoured the latter for this zeroes treatment : the 
complex action problem above being amenable to treatment via quantum fluctuations. The 
equivalence of the quasi-classical approach to the continuum is only resolved when 
both spatial and time extent limits are simultaneously reached, but more over the 
symmetry breaking mechanisms are otherwise potentially contaminated with disconnected 
contributions to the partition function. Conversely, the nonlocal nature of the 
intermediate result of Euclidean time lattice spacing for the factored product case 
means that nonlocal symmetry breaking terms can be considered but on a countably 
finite space, yielding well-defined thermodynamic functions in the continuum. The 
quantum fluctuations represented through the partition function are inherently 
nonlocal and it is only through this latter limiting formalism that the disconnected 
contributions are exactly articulated since otherwise, as we have seen, the generating 
functional for the partition function becomes simply a generating function.

There is an intrinsic connection between this discussion and the renormalization 
group arguments made by Wilson through the implicit differentiability conditions 
introduced for the spatial lattice volume extent~\cite{ball}. 
Looking at the density relation, 
the condition used to establish the $N\!\rightarrow\!\infty$ zeroes singularities, 
it is possible to understand that the critical points on the real axis of the 
expansion parameter correspond to the fixed points of the renormalization group 
equation defined in the limit $N\!\rightarrow\!\infty$. This is a new result of the 
loop operator treatment since for the general classical system partition function 
zeroes case it is not possible to make this association, as the asymptotic scaling 
properties of the expansion parameter are not defined. The expansion parameter in this 
case can be strongly dependent on the Euclidean time lattice spacing and singular and 
dependent on the asymptotic properties therein. 
The differentiability conditions of the renormalization 
group equations have an obvious expression in the recursion relation for the 
Laplace transforms we have discussed, 
and the group property can be understood in this mapping between 
successive poles. Returning to the Vafa-Witten comment it is of course possible as with 
the renormalization group discussion that discontinuities between minima of the 
vacua arise in the asymptotic limit, in which case it is possible to discuss phase 
transitions even in the case that we have no intensive source term comparing with the 
Ising model with external field~\cite{vafa}. 
The difference in level between the peaks of the 
vacua essentially then does not have to be infinite in order that an exact 
discontinuity arises in the asymptotic limit of the expansion parameter $\beta$ 
which would 
now play an analogous role to time in the motivation of the renormalization group 
equations. 

In this loop operator treatment it is therefore important to understand 
the evolution of the spatial lattice spacing dependence in regards to the nonlocal 
contributions to the expansion : infrared divergences potentially lie in the 
nonlocality of the expansion. It is important to understand as well how the 
singularities of the expansion relate to symmetry breaking and the emergence of 
Goldstone modes which we will now discuss.

\section{Asymptotic Freedom}
The properties of branch cuts are clearly important in the partition function zeroes 
singularity picture as they relate to the emergence of a form of
geometric bound on the modulus and argument of the zeroes in the complex expansion 
parameter plane~\cite{lee+yang2}. 
It is only the property of the zero nearest the real axis which is 
strictly important in determining the singularities in the $N\!\rightarrow\!\infty$
limit, but in the quantum case this also establishes the behaviour of an asymptotically 
defined source for the singularity. The scaling of the feature is therefore 
critically important to determining whether or not singularities of the zeroes in 
the spatial lattice polynomial expansion correspond to fixed points of this 
renormalization group equation defined along the real axis in $\beta$.
The sense in which we associate the loop operator partition function zeroes 
treatment to the renormalization group equation approach is perhaps therefore closer 
to discussions of the strong interaction and a momenta dependent 
treatment~\cite{strong}. Although 
the spatial volume is clearly important we are not treating the lattice spacing 
dependencies of this extent in a direct manner, and so the $N\!\rightarrow\!\infty$ 
properties of the zeroes are in a sense disconnected from the asymptotic treatment 
of the expansion parameter : in the sense that we would be interested to relate the 
correlation length and bare parameters of the system to finite $N$ scaling properties 
of the volume. 

Never the less the symmetry breaking mechanisms of the system can be 
understood tentatively with relation to the Euclidean time extent and correlation 
length.
If we take a conventional renormalization group analysis of a system like the 2-d 
$O(3)$ model we can understand that the correlation length, relating to the 
conventional renormalization group $\beta$-function coefficients, diverges 
exponentially in the limit of vanishing lattice spacing 
and so there are massless particles~\cite{Dtheory}. 
However, for this particular system, 
the Mermin-Wagner-Coleman theorem prevents the massless Goldstone 
bosons interacting and so the model has a nonperturbatively generated mass 
gap~\cite{MWC}. 
Conversely other systems can be pictured in which the correlation length remains 
infinite and we have a broken symmetry, but reducing the Euclidean time extent $\beta$ 
eventually recovers a symmetric phase with a finite correlation length. If the 
Euclidean time extent is very much smaller than the correlation length we can consider 
that all these systems undergo dimensional reduction : we can treat one direction as 
effectively static. This then allows the continuum limit of these models to be 
defined with $\beta\!\rightarrow\!\infty$. This limit is of interest because we are 
interested in the asymptotic properties of $\beta$ : which fixed point, or 
singularity on the real axis, we end up at in this limit.
Notice that we have not aimed to fix at any point an association of the loop operator 
formalism with this dimensional reduction. In essence we therefore invert the above 
argument : having determined singularities in the limit $N\!\rightarrow\!\infty$ we 
then consider the properties of these points on the real $\beta$ axis where asymptotic 
convergence to the continuum is in principle guaranteed by construction. This is 
particularly useful form for the direct analysis of numerical lattice results since 
it is possible to understand from direct examination of the singularities whether or 
not they correspond to genuine fixed points of a given model or are simply lattice 
spacing dependent cutoff effects.

Considering the arguments defining a renormalization program for QED to all 
orders perturbatively in terms of momenta, one must at some point make momentum 
cutoff dependent subtractions in the diagrams~\cite{strong}. 
Prescriptively constants are chosen so that the relevant Ward identities are 
satisfied at one particular value of momentum, and such that the 
renormalization functions do not contain singularities for the zero electron mass case. 
Similarly by construction the renormalized amplitudes of the diagrams do not 
contain infrared divergences for zero photon mass and nonzero electron mass. 
This prescription for infrared divergences of Gell-mann and Low's discussed 
corresponds to the 
asymptotic limit of momentum cutoff being much bigger than the mass, and 
the solutions of the renormalization group equations for all momentum cutoff values 
and zero photon mass necessarily then define a scale invariant field theory.
One can break this scale invariance in one of two ways ; by letting the mass 
be nonzero or by choosing a solution of the renormalization group equations at zero 
mass that is not a fixed point.
One can therefore understand by analogy that for the 
loop operator approach all singularities in the spatial $N$ extent can be connected 
to a finite $\beta$ value with an asymptotic form, but that these points don't 
necessarily correspond to fixed point solutions if they do not have any support. 
To 
elucidate slightly : if a zero has an edge singularity associated with boundedness 
then its form is akin to this 
former case and we would have a broken symmetry with a correlation length that 
diverges as the continuum limit is approached, 
and the converse is that of a finite correlation length with a valid continuum limit.
The third case is that of the mass gap, which would be otherwise under-defined in the 
analysis of say the case for example of the Ising model with external field, 
since some points of the 
complex phase plane diagram are necessarily undefined with the Laplace transform of the 
poles having essentially an under-defined contour~\cite{vafa}.  

Consider as an example the Haldane 
conjectured antiferromagnetic $O(3)$ quantum spin chains, where 
integer spins have a mass gap and half-integer spins are gapless~\cite{Dtheory}.
The low-energy effective theory for this system is the 2-d $O(3)$ model at 
vacuum angle $\theta = \pi$, which is in the universality class of the $k=1$ 
Wess-Zumino-Witten model. Numerical studies of this model indicate that the gap which is 
present for all $\theta \neq \pi$, vanishes at the specific value $\theta = \pi$. When the 
Euclidean time extent $\beta$ is made finite this topological term also disappears. 
One can conclude from this system 
therefore that one would expect to see a singularity on the $\beta$ axis from the 
loop operator formalism, but that this should not scale directly to the continuum because 
the finite volume dependence of the scheme essentially introduces a divergent correlation 
length and 
one sees otherwise a pseudo-critical scaling. 
However the lattice singularity still exists on the 
finite volume and its relation to the asymptotic scaling can be measured directly from 
the lattice result, since again one can obviously study the gap dependence on finite 
system size directly in numerical simulation. The advantage with the new loop operator 
method is firstly that we can quantify the directly the asymptotic scaling properties 
of the gap itself, but also understand the evolution of other cutoff dependent quantum 
phase transition effects from the scaling arguments of related singularities. 

Preliminary results for the loop operator zeroes 
scheme are presented for the case of mixed integer - 
half-integer spin chains which share a related gapless property to the integer chains 
but with an additional saddle point effect~\cite{n=4}. 
These admit a natural probabilistic loop 
operator treatment as well following the pedagogical S=1/2 AFM 
treatment schemes we have followed and discussed. 
An exact treatment has been recently given for the dimensional reduction 
aspects of plaquette defined operators over an extended phase space for the case of 
gauge field theories. This raises then raises the possibility of performing a similar 
analysis to the loop operator treatment for the case of gauge fields, as opposed to 
quasi-classical treatment, but the issues of numerically simulation are by no means 
trivial to address. A general pathology of the factorised loop-cluster method, where 
the Monte Carlo decisions are taken plaquette wise, is the statistical limitations of 
building up a long tailed probability distribution, and extending to the case of gauge 
or complex valued fields one could then have also non positive definite operator elements which 
do not yield a probabilistic interpretation at all. One option is to use a projection 
operator basis acting on a conventional globally updated space to define pseudo-plaquette
states~\cite{su2a}\cite{su2b}. 
Strictly from the Laplace transform arguments we have reviewed and presented 
the only necessary criterion for a plaquette term on the space is that it contain at 
most one pole and have an continuous probability density function. In QCD, for example, 
a basis of related operators can be compiled over a given lattice gauge field ensemble 
such that the poles of the basis suppress unphysical state contributions
through a natural ranking~\cite{N*}. Here we define first a Euclidean time dependent polynomial 
form for the global update scheme, independent of the related operator basis, and introduce a 
limiting product such that the elements of these pseudo-plaquette 
operators contain at most one zero of the polynomial, which is easily done since the 
probability density functions of the operators on the globally generated gauge ensemble 
become universally distributed at some critical scale~\cite{immu}. 
One can understand this latter result from 
general consideration of the central limit theorem in regard to defining the contour 
for the above Laplace transforms, and its relation to the stationary phase approximation in 
this regard~\cite{Deltaf}.

In lattice field theory calculations one can avoid some of 
the uncertainties associated with truncating a perturbative expansion in determining 
constants in the renormalization group equations by defining various nonperturbative 
schemes~\cite{rimoma}\cite{rimomb}. 
Constants are set through the chiral Ward identities for the lattice Green 
functions. For example with RI/MOM the conditions are set on Green functions at 
large external momenta, while in the Schr\"odinger Functional scheme they are set in small 
lattice volumes. The lattice renormalization scales have to be larger than 
$\Lambda_{QCD}$, to connect with perturbation theory and smaller than the 
inverse lattice spacing to avoid cutoff effects, which is perhaps 
computationally expensive. With the RI/MOM approach as well the pseudoscalar Green 
function, which couples to the Goldstone boson, has a leading contribution 
diverging in the chiral limit. 
Similarly in building candidate supersymmetric lattice models one is interested 
in the Ward identities in the continuum limit in terms of renormalized
perturbation theory~\cite{susy_rev}. Schemes involve, to finite order in perturbation theory, 
adjusting a coefficient in local terms in the action but here as well it is possible 
for the action to become complex~\cite{susy_ur}. In both cases it is possible to envisage 
calculating a simple limiting operator factorisation with some Euclidean time extent 
polynomial 
scheme defined such that the asymptotic properties of these lattice renormalization 
properties could be further studied directly in this context.

\section{Summary}
We have discussed, in the context of recent ideas on the r${\rm\hat{o}}$le of dimensional reduction 
in quantum systems and reviews of loop operator methods a formulation of a partition 
function polynomial zeroes treatment for quantum systems. We have argued that in general the 
quantum case should be treated separately from the classical case through a local projection 
operator argument for the factored product, suggesting the evolution of infrared divergences is 
otherwise implicit to the quantum case. We have relaxed the condition that the expansion variable 
be expressed through an exponential function to this end, whilst still retaining its asymptotic structure.
By working over an extended phase space we have then identified a limiting 
form with a connection to the continuum, and the fixed point solutions of renormalisation group 
equations, using recent results for the treatment of exact probabilistic lattice 
dynamics. Related to this 
continuum limiting process we have then discussed the prospects for further evaluating gauge field 
theories for the purposes of perturbation theory independent renormalization group analysis, and the 
treatment of theories with a non positive-definite source terms in the Euclidean action. 
The outline is necessarily general 
but further investigation of the limiting analyticity of the partition function is proposed, with 
three cases with relevance to symmetry breaking mechanisms envisaged as pertinent for future analysis 
and categorisation.

\end{document}